\documentclass[prl,twocolumn,showpacs,superscriptaddress]{revtex4}

\usepackage{epsfig}
\usepackage{amssymb}
\usepackage{graphicx, natbib}
\begin{document}

\title{Role of relaxation in the quantum measurement of a superconducting qubit using a nonlinear oscillator}

\author{T. Picot}
\affiliation{Kavli Institute of Nanoscience, Delft University of Technology, PO Box 5046, 2600GA Delft, The Netherlands}
\author{A. Lupa\c scu}
\affiliation{present adress: Laboratoire Kastler Brossel, ENS - 24 rue Lhomond, 75005 Paris, France}
\author{S. Saito}
\affiliation{NTT Basic Research Laboratories, NTT Corporation, 3-1 Morinosato-Wakamiya, Atsugi-shi, 243-0198, Japan }
\author{C.J.P.M. Harmans}
\affiliation{Kavli Institute of Nanoscience, Delft University of Technology, PO Box 5046, 2600GA Delft, The Netherlands}
\author{J.E. Mooij}
\affiliation{Kavli Institute of Nanoscience, Delft University of Technology, PO Box 5046, 2600GA Delft, The Netherlands}

\date{ \today}

\begin{abstract}
We analyze the relaxation of a superconducting flux qubit during measurement. The qubit state is measured with a nonlinear oscillator driven across
the threshold of bifurcation, acting as a switching dispersive detector. This readout scheme is of quantum non-demolition type. Two successive
readouts are used to analyze the evolution of the qubit and the detector during the measurement. We introduce a simple transition rate model to
characterize the qubit relaxation and the detector switching process. Corrected for qubit relaxation the readout fidelity is at least 95\%. Qubit
relaxation strongly depends on the driving strength and the state of the oscillator.
\end{abstract}

 \maketitle


  Superconducting qubits are quantum systems based on microfabricated superconducting circuits with one or more Josephson junctions as nonlinear elements
\cite{Devoret-review}. They are artificial quantum systems, with properties that can be defined by design of the mesoscopic parameters of the
circuit. In superconducting qubits quantum state readout is of considerable interest, since the fabricated nature of qubit and detector allows full
control of the qubit-detector coupling strength. Consequently, aspects of quantum measurement can be experimentally investigated that are commonly
not easy to access. These include the realization of high fidelity \cite{Lupascu-high-contrast} and projective \cite{Adrian-QND,Boulant}
measurements, partial measurements \cite{Katz}, and the continuous observation of qubit dynamics \cite{Korotkov,Ilitchev}. In addition, state readout
is a subject relevant to quantum computing: projective measurements are an essential part of protocols for quantum information processing.

Qubit state readout can be performed in various ways. In dispersive readout the qubit is coupled to an oscillator, with a quadratic type of
interaction. As a result of this nonlinear coupling, the resonance frequency of the oscillator becomes qubit-state dependent
\cite{Ilitchev,Lupascu-first-paper}. The state of the qubit can thus be inferred from a measurement of the properties of the oscillator.


Nonlinear switching detectors are very attractive as they are able to amplify the information extracted from the qubit, leading to very fast readout
with high fidelity. Here we present a detailed experimental analysis of switching dispersive readout of a superconducting flux qubit with a nonlinear
oscillator. We introduce a simple model that allows to characterize the detector switching process and the qubit relaxation induced by the operation
of the detector. We find that the main source of measurement error is qubit relaxation induced by the operation of the detector.


\begin{figure}[!]
\includegraphics[width=3.4in]{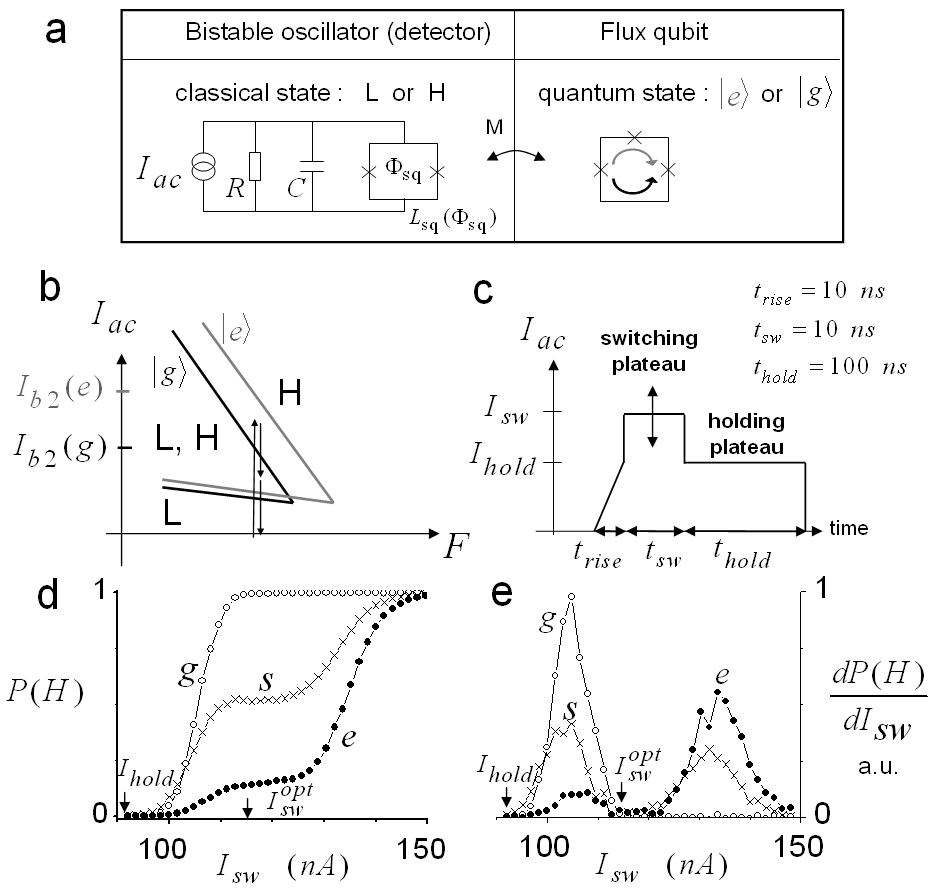}
\caption{\label{dispersive-squid} (a) Qubit and readout circuit. (b) Bistability diagram of the oscillator for the two
qubit states $|g>$ or $|e>$. (c) Oscillator driving amplitude for qubit readout. (d) Oscillator switching probability, $P(H)$, as a function
of the switching plateau driving amplitude $I_{sw}$ for the qubit prepared initially with a Rabi pulse, $0(g)$, $\pi(e)$ , $3\pi/2(s)$.
(e) First derivative of $P(H)(I_{sw})$.}
\end{figure}

\begin{figure}[t]
\includegraphics[width=3.4in]{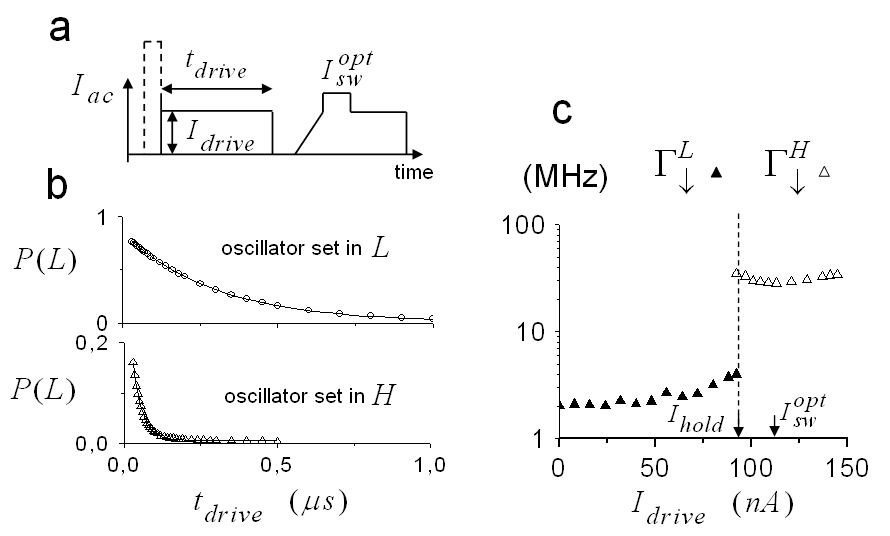} \caption{\label{relaxation} (a) Oscillator driving sequence used to analyze the qubit relaxation.
In the first driving pulse the oscillator is set either in $L$ or $H$ (see text). (b) Decay of the probability for the measurement outcome $L$ (due to the
qubit relaxation). (c) Qubit relaxation rates versus the oscillator driving amplitude $I_{drive}$ for the oscillator in the state $L$ or $H$.}
\end{figure}

  The persistent current flux qubit circuit \cite{Hans} is a superconducting loop interrupted by three Josephson junctions (see figure
\ref{dispersive-squid}a). Biased with an external flux $\Phi_{qb}$ close to half a flux quantum $\Phi_0/2$, it behaves as a quantum two-level system.
The ground state $|g>$ and the excited state $|e>$ are quantum superpositions of two oppositely circulating persistent currents ($\pm I_p$). In the
basis of the current operator $\hat I = I_p \hat \sigma_z$, the Hamiltonian of the flux qubit is: $\hat H= \frac{1}{2}(\epsilon \hat \sigma_z+\Delta
\hat \sigma_x)$, where $\epsilon=2 I_p(\Phi_{qb}-\frac{1}{2}\Phi_0)$ and $\Delta$ is the quantum tunneling energy between the two current states.
Here $\Delta/h=5$ GHz and the qubit is operated at a frequency $F_{qb}=\sqrt{\epsilon^2+\Delta^2}/h=14.2$ GHz.

  Our detector is an oscillator formed by the inductance of a DC-SQUID and a capacitor. The inductance of the DC-SQUID and consequently the resonance
frequency of the oscillator $F_0$ depend on the flux $\Phi_{sq}$ enclosed in the SQUID loop. The oscillator is operated at a frequency $F_0=1.5$ GHz
and has a quality factor $Q=20$. The SQUID inductance is nonlinearly dependent on the SQUID current. Due to this nonlinearity, when the oscillator is
driven at a frequency $F<F_0(1-\frac{\sqrt{3}}{2Q})$, it can switch between a state of low oscillation amplitude (labeled $L$) and a state of high
oscillation amplitude (labeled $H$) \cite{Siddiqi}. Three situations are possible depending on the amplitude of the driving current $I_{ac}$ compared
to the two bifurcation currents $I_{b1}(F)<I_{b2}(F)$. For weak driving $I_{ac}<I_{b1}$ the oscillator is in the $L$ state. For strong driving
$I_{ac}>I_{b2}$ the oscillator is in the $H$ state. For intermediate driving, $I_{b1}<I_{ac}<I_{b2}$ the oscillator is bistable and can be in either
of the $L,H$ states.

  Due to the flux-dependence of the SQUID inductance, the upper bifurcation current $I_{b2}$ is highly sensitive to the flux $\Phi_{sq}$. As
the expectation value of the flux generated by the qubit is different for the two energy eigenstates, $I_{b2}$ depends on the qubit state. In
particular we choose the qubit bias such that $I_{b2}(g)<I_{b2}(e)$. For the readout of the flux qubit, the oscillator driving amplitude $I_{sw}$ is
increased to a value $I_{b2}(g)<I_{sw}<I_{b2}(e)$ (see figures \ref{dispersive-squid}b and \ref{dispersive-squid}c), such that the oscillator
switches to the $H$ state with a high probability if the qubit is in $|g>$, while it stays in the $L$ state if the qubit is $|e>$. This first time
interval (the switching plateau) with duration $t_{sw}$ constitutes the actual measurement interaction. To optimally discriminate between the two
oscillator states $L$ and $H$, noise from the detection electronics needs to be averaged out. This is performed during the holding plateau
($t_{hold}$) with the driving amplitude decreased such that both oscillator states can be maintained without switching or retrapping.

 It should be noted that the large difference between $F_{qb}$ and $F_0$ inhibits energy exchange between the qubit and the oscillator. The states
$|e>$ and $|g>$ are thus preserved during the measurement.

As shown in figures \ref{dispersive-squid}d and \ref{dispersive-squid}e, the two bifurcation currents $I_{b2}(g)$ and $I_{b2}(e)$ can be remarkably
well resolved. The qubit readout is performed at an amplitude $I^{opt}_{sw}=115$ nA, where the switching probability is $P^g(H)=99.7\%$ if the qubit
is in $|g>$ and $P^e(H)=14.6\%$ if the qubit is in $|e>$, resulting in a readout contrast of 85\%. The main loss of readout contrast occurs when the
qubit is in $|e>$, suggesting that the readout fidelity is limited by the qubit relaxation before or during the measurement.

  In practical flux qubits one finds a very irregular dependence of the relaxation on flux bias due to electromagnetic modes and 'natural' two-level
systems. The qubit is operated at a flux bias where the relaxation rate is locally minimal. During measurement, the SQUID transport current varies in
time. By second-order processes, this oscillation shifts the average value of the circulating current thus shifting the qubit bias. At that new
point, relaxation is likely to be faster. More seriously, the qubit flux bias is swept at the oscillator driving frequency. In the $H$ state, the
flux is modulated over a range as large as 5 m$\Phi_0$ corresponding to a sweep of the energy splitting over 5 GHz. The qubit can thus be swept
through regions where the relaxation rate is much higher. Moreover, when the oscillator is driven into its nonlinear regime new channels of
relaxation might open where qubit energy is transferred directly to the oscillator \cite{Ioana-relaxation}.

  We first measure the qubit relaxation under conditions where the oscillator is fixed in either the $L$ or the $H$ state. The qubit is initially prepared in $|e>$. Next the oscillator is set either in the $H$ state or in the $L$ state. To prepare $H$ a short high
driving pulse $I_{ac}>I^e_{b2}$ (dashed line in figure \ref{relaxation}a) is applied. Subsequently the oscillator is driven for a time $t_{drive}$ at
an amplitude $I_{drive}$. Afterwards the qubit state is read out with a regular measurement pulse. Figure \ref{relaxation}b as an example gives the
decay in time of the probability for readout in the $L$ state when the oscillator is driven with an amplitude $I_{drive}=I_{hold}$. This decay is
exponential and is due to qubit relaxation. The two qubit relaxation rates $\Gamma^L_{\downarrow}$ and $\Gamma^H_{\downarrow}$ for the two states of
the oscillator are significantly different. Figure \ref{relaxation}c shows the dependence of the qubit relaxation rates on the oscillator driving
amplitude for $I_{drive} < I_{hold}$ and $I_{drive} > I_{hold}$ when the oscillator is in the $L$ or $H$ state, respectively. Whereas
$\Gamma^H_{\downarrow}$ is almost constant, $\Gamma^L_{\downarrow}$ increases with the driving amplitude.

Similarly, we characterize the qubit relaxation during the rising part of the readout pulse with the effective relaxation rate
$\Gamma^{rise}_{\downarrow}$ obtained from an exponential decay fitting of $P(L)$ as a function of the rise time $t_{rise}$. Corrected for the
relaxation during the rising part of readout pulse (5\%) and initial qubit preparation errors (5\%)
\cite{info-prep}, the readout fidelity is $f=95\%$. The remaining errors occur during the switching plateau.

 The approach used so far is well suited to analyze the qubit relaxation for a driving amplitude where the oscillator is in a stable state.
However for higher driving amplitude $I_{drive}>I_{hold}$ and especially at the driving amplitude during the switching plateau, the $L$ state is
metastable and can switch to the $H$ state. Therefore to analyze the qubit relaxation, the oscillator switching process needs to be included.

\begin{figure}[!]
\includegraphics[width=3.4in]{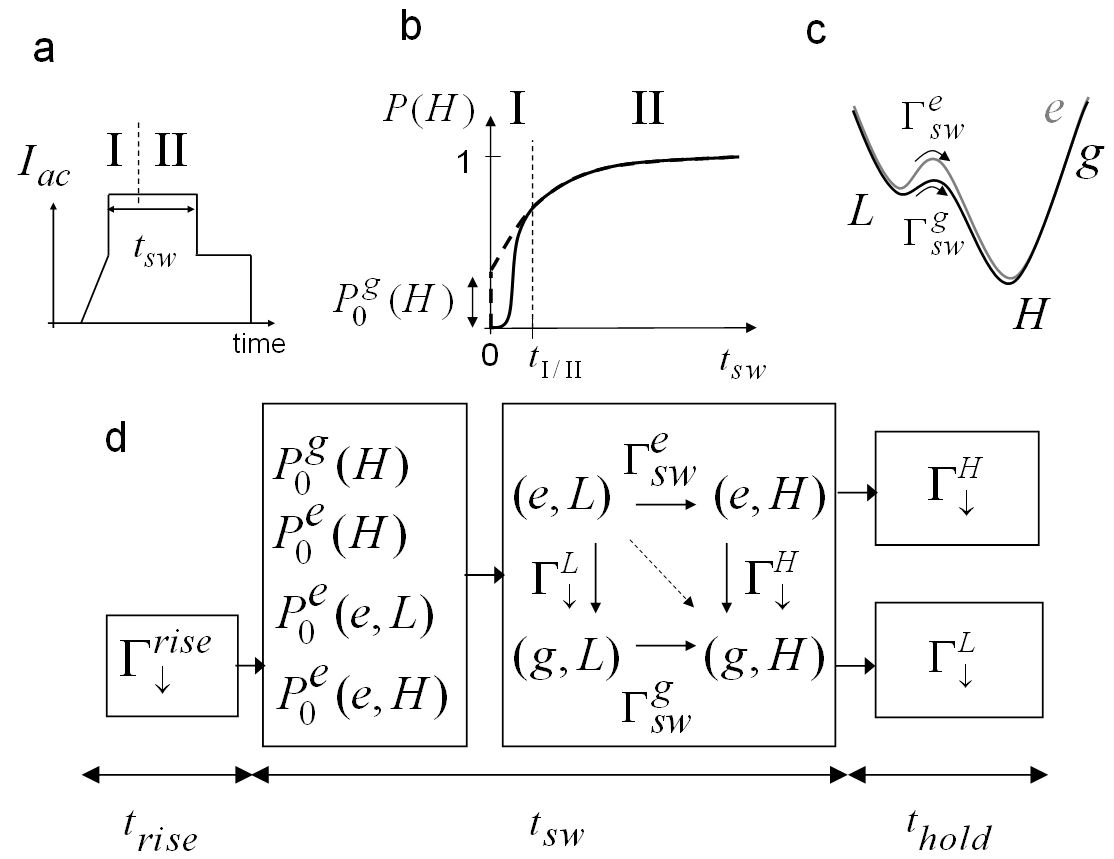} \caption{\label{model-all-parameters} (a) Oscillator driving amplitude.
(b) Switching probability as a function of the duration of the switching plateau (solid line). (c) Effective potential barrier between $L$ and $H$, for
the qubit state $|g>$ or $|e>$. (d) Schematic of the model used to describe the evolution of the qubit and the oscillator during the readout pulse,
including qubit relaxation rates $\Gamma^L_{\downarrow},\Gamma^H_{\downarrow}$ and oscillator switching rates $\Gamma^e_{sw},\Gamma^g_{sw}$.
$P^g_0(H),P^e_0(H),P^e_0(e,H),P^e_0(e,L)$ describe the regime I (see text).}
\end{figure}

In the following we analyze the oscillator switching and the qubit relaxation during the switching plateau. The solid line in figure
\ref{model-all-parameters}b shows the oscillator switching probability when the qubit is in $|g>$ as a function of the duration of the switching
plateau $t_{sw}$. We distinguish two regimes, indicated as I and II. The boundary of the two regimes is after a time of about 10 ns, indicated as
$t_{I/II}$. In regime I the switching probability increases very fast. At the origin of the switching plateau, the oscillator driving amplitude is
increased in about 1 ns, probably leading to non-adiabatic effects. In regime II the switching probability increases with a constant rate. The switching rate $\Gamma_{sw}$ from the $L$ state to the $H$ state depends on the height of the effective potential barrier
\cite{Dykman-Rates} $\Delta U=U_0 [1-(I_{ac}/I_{b2})^2]^{3/2}$ between $L$ and $H$. For a driving current $I_{ac}$ close to the upper bifurcation
current $I_{b2}$, the oscillator switching rate $\Gamma_{sw}$ increases strongly by a few orders of magnitude. As $I_{b2}$ is different for the two
qubit states, $\Gamma_{sw}$ strongly depends on the qubit state, which is the principle of the measurement. As $I_{b2}(g)< I_{b2}(e)$, it follows
that $\Gamma^g_{sw}<<\Gamma^e_{sw}$ (figure \ref{model-all-parameters}c).

We describe the qubit and the oscillator as a 4-state system $(g,L)$, $(g,H)$, $(e,L)$ and $(e,H)$, with 4 corresponding occupation probabilities.
Due to normalization, only 3 probabilities are independent. We use $P(H)=P(e,H)+P(g,H)$, $P(e,H)$ and $P(e,L)$. In regime II, oscillator switching
($L \rightarrow H$) and qubit relaxation ($e \rightarrow g$) are described with the set of four rates $\Gamma^g_{sw},\Gamma^e_{sw}$ and
$\Gamma^L_\downarrow, \Gamma^H_\downarrow$ (figure \ref{model-all-parameters}d). Oscillator retrapping ($H \rightarrow L$) and qubit excitation ($g
\rightarrow e$) are negligible. The evolution of the occupation probabilities is given by:

\begin{equation}
\label{differential-equations}
\left\{
\begin{array}{lll}
\frac{dP(e,L)}{dt} & = & - P(e,L) ( \Gamma^e_{sw} + \Gamma^L_{\downarrow} ) \\
\frac{dP(e,H)}{dt} & = & P(e,L) \Gamma^e_{sw} - P(e,H) \Gamma^H_{\downarrow} \\
\frac{dP(g,L)}{dt} & = & P(e,L) \Gamma^L_{\downarrow} - P(g,L) \Gamma^g_{sw}  \\
\frac{dP(g,H)}{dt} & = & P(g,L) \Gamma^g_{sw} + P(e,H) \Gamma^H_{\downarrow} \\
\end{array}
\right.
\end{equation}

  For a given initial qubit state $i$ ($g$ or $e$) at the origin of the switching plateau $t=0$, we denote $P^i_0(H)$, $P^i_0(e,H)$, and $P^i_0(e,L)$ as
the initial conditions for the set of equations (\ref{differential-equations}). They quantify the evolution of the qubit and the oscillator evolution
during regime I. We choose to define the initial conditions at $t=0$, and not at $t=t_{I/II}$. They are obtained by extrapolating $P(H)$, $P(e,H)$
and $P(g,L)$ in regime II to $t=0$. In the more general case of an initial qubit state with an occupation probability of $|e>$: $P(e)=x$, the initial
conditions are: $P_0(H)=x P^e_0(H)+ (1-x) P^g_0(H)$ (and similarly for $P_0(e,H)$ and $P_0(e,L)$).

  If the qubit is initially in $|g>$, the only relevant rate is $\Gamma^g_{sw}$ and the set of equations (\ref{differential-equations})
reduces to: $dP(H)/dt=[1-P(H)]\Gamma^g_{sw}$. For a given switching plateau driving amplitude $I_{sw}$, $P^g_0(H)$ and $\Gamma^g_{sw}$ are extracted
by fitting the switching probability $P(H)$ as a function of $t_{sw}$ with $P(H)=1-[1-P^g_0(H)]\exp{(-t_{sw}\Gamma^g_{sw})}$ (as shown by the dashed
line in figure \ref{model-all-parameters}b).

  The driving amplitude $I^{opt}_{sw}$ used for the measurement is higher than $I_{b2}(g)$. In that case the $L$ state does not exist when the
qubit is in $|g>$, so qubit relaxation directly causes oscillator switching. The state $(g,L)$ is eliminated from the set of equations
(\ref{differential-equations}) by assuming that $\Gamma^L_{\downarrow}$ corresponds to the transition $(e,L) \rightarrow (g,H)$ shown by the dashed
arrow in figure \ref{model-all-parameters}d. If the qubit is in $|e>$, two switching processes are possible, either due to qubit relaxation (rate
$\Gamma^L_{\downarrow}$), or to switching while the qubit is excited (rate $\Gamma^e_{sw})$. The sum
$\Gamma_{sw}=\Gamma^L_{\downarrow}+\Gamma^e_{sw}$ and $P^e_0(H)$ can be extracted from $P(H)$, given by: $P(H)=1-[1-x
P^e_0(H)]\exp{(-t_{sw}\Gamma_{sw})}$, where $x$ is the occupation probability of $|e>$ at the origin of the switching plateau \cite{info-x}.
Depending on the process, the state of the qubit after switching is either $|g>$ or $|e>$. Therefore, it is possible to discriminate between the two
processes by measuring the qubit state after a switching event. At the end of the switching plateau $P(e,H)$ is given by:

\begin{figure}[!]
\includegraphics[width=3.4in]{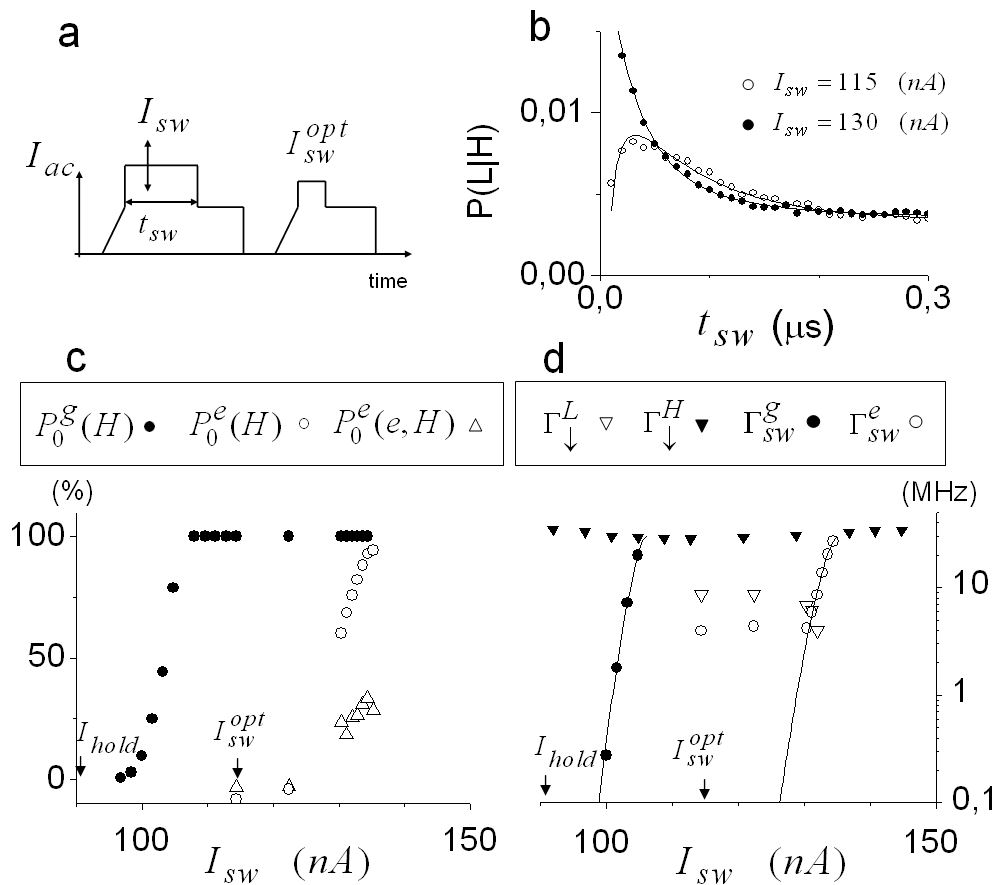} \caption{\label{conditional-switching} (a) Sequence of two successive readouts used to analyze
the qubit state after the first measurement. (b) Conditional probability for the measurement outcome $L$ after a switching event in the first readout.
(c) $P^g_0(H), P^e_0(H), P^e_0(e,H)$ and (d) oscillator switching rates $\Gamma^e_{sw}, \Gamma^g_{sw}$ and qubit relaxation rates
$\Gamma^L_{\downarrow}, \Gamma^H_{\downarrow}$ as a function of the switching plateau driving amplitude $I_{sw}$.}
\end{figure}

\begin{equation}
\label{qubit-after-switching}
\begin{array}{l}
P(e,H)(t_{sw})  = x P^e_0(e,H) e^{-t_{sw} \Gamma^H_{\downarrow}} \\
    \qquad + [1 - x P^e_0(H)]\frac{\Gamma^e_{sw}}{\Gamma^H_{\downarrow}-\Gamma_{sw}}\big[ e^{-t_{sw}\Gamma_{sw}}-e^{-t_{sw}\Gamma^H_{\downarrow}} \big]
\end{array}
\end{equation}

 If relaxation is the only process which can lead to oscillator switching then $P(e,H)=0$. In equation (\ref{qubit-after-switching}), the first term
proportional to $P^e_0(e,H)$ corresponds to switching events during regime I, whereas the second term proportional to $\Gamma^e_{sw}$ corresponds to
switching events during regime II.

 The occupation probability of $|e>$ after a switching event, given by the conditional probability $P(e|H)=P(e,H)/P(H)$, is measured with a
second, successive readout. The conditional probability $P(L|H)$ for a measurement outcome $L$ after a switching event is $P(L|H)=A P(e|H) + B$,
where $A=0.9\exp{(-t_{hold}\Gamma^H_{\downarrow})}$ includes the contrast of the second readout and the qubit relaxation, and $B=3.5$ $10^{-3}$ is
the finite measurement error when the qubit is in $|g>$. When the oscillator is in the $H$ state, qubit relaxation during the holding plateau is very
strong. Therefore, the remaining signal is very small $P(L|H) \approx 10^{-2}$. A key point for our analysis is the very high contrast of the readout
allowing a resolution of $P(L|H)$ as small as $10^{-3}$. $\Gamma^e_{sw}$ and $P^e_0(e,H)$ are extracted from a fit of $P(e,H)$ to equation
(\ref{qubit-after-switching}). The decay rates $\Gamma_{sw}$ and $\Gamma^H_{\downarrow}$ are fixed (extracted previously), and $P^e_0(e,H)$ and
$\Gamma^e_{sw}$ are fitting parameters. The obtained values of $P^e_0(e,H)$, $\Gamma^e_{sw}$ and $\Gamma^L_{\downarrow}$ are shown in figures
\ref{conditional-switching}c,d. For comparison, $P^g_0(H)$, $P^e_0(H)$, $\Gamma^g_{sw}$ and $\Gamma^H_{\downarrow}$ obtained previously are shown as
well.

  The ratio $\Gamma^e_{sw}/\Gamma^g_{sw}$ determines the intrinsic readout fidelity. The solid black lines in figure
\ref{conditional-switching}d are fits of $\Gamma^e_{sw}$ and $\Gamma^g_{sw}$ using the oscillator escape rate equation as given in
\cite{Dykman-Rates}, from which we extract $I_{b2}(g)=108$ nA and $I_{b2}(e)=136$ nA. Extrapolating $\Gamma^e_{sw}$, the intrinsic readout fidelity
would be 99.9 \%. However, in practice we observe a saturation of $\Gamma^e_{sw}$ at approximately 4 MHz.

The qubit relaxation rate $\Gamma^L_{\downarrow}$ can directly limit the readout fidelity. The independent determination of $\Gamma^L_{\downarrow}$
and $\Gamma^e_{sw}$ is an important result of this paper. At the driving amplitude $I^{opt}_{sw}$ used for the measurement, $\Gamma^L_{\downarrow}$
is higher than $\Gamma^e_{sw}$. Therefore, for measurement times beyond the oscillator transient period, qubit relaxation is the main process
limiting the readout fidelity.

To summarize, we have introduced a simple model to characterize the qubit relaxation and the oscillator switching process during measurement of a
flux qubit. Qubit relaxation increases significantly with increasing driving strength with the oscillator in the low-amplitude state; it jumps to a
much higher rate when the oscillator switches to its high-amplitude state. Corrected for qubit relaxation and initial qubit preparation errors, the
readout fidelity is at least 95 \%.

  We thank F. Wilhelm, I. \c Serban, S. Ashhab, Y. Nazarov and H. Wei for useful discussions. This work was supported by the Dutch Organization
for Fundamental Research on Matter (FOM), E.U. EuroSQIP, the E.U. Marie Curie program, and the NanoNed program.


\begin{thebibliography}{20}
\bibitem{info-prep}
  Taking into account the period $T=3.2$ ns and the decay time $\tau=15$ ns of the Rabi oscillations, we estimate the preparation fidelity of $|e>$ to
  be 95\%.
\end{thebibliography}
\end{document}